\documentclass[twocolumn]{aastex631}

\usepackage{ulem}
\usepackage{enumitem}
\usepackage{lineno}

\newcommand{\ergps}{$\mathrm{erg~s^{-1}}$}

\newcommand{\swift}{Swift~J1818.0$-$1607}
\newcommand{\pcpcc}{$\rm pc~cm^{-3}$}
\newcommand{\maspy}{$\rm mas~yr^{-1}$}
\newcommand{\kmps}{$\rm km~s^{-1}$}
\newcommand{\mjypb}{$\rm mJy~beam^{-1}$}

\newcommand{\multilinecomment}[1]{}

\shorttitle{VLBA astrometry of \swift}
\shortauthors{Ding et al.}
\graphicspath{{./}{Figures/}}

\begin{document}

\title{VLBA Astrometry of the Fastest-spinning Magnetar \swift:\\ A Large Trigonometric Distance \& A Small Transverse Velocity}

\author[0000-0002-9174-638X]{Hao Ding}
\altaffiliation{EACOA Fellow}
\affiliation{Mizusawa VLBI Observatory, National Astronomical Observatory of Japan, 2-12 Hoshigaoka, Mizusawa, Oshu, Iwate 023-0861, Japan}
\email{hdingastro@hotmail.com}

\author[0000-0001-9208-0009]{Marcus E. Lower}
\affiliation{Australia Telescope National Facility, CSIRO, Space and Astronomy, PO Box 76, Epping, NSW 1710, Australia}

\author[0000-0001-9434-3837]{Adam T. Deller}
\affiliation{Centre for Astrophysics and Supercomputing, Swinburne University of Technology, John St., Hawthorn, VIC 3122, Australia}
\affiliation{ARC Centre of Excellence for Gravitational Wave Discovery (OzGrav), Australia}

\author[0000-0002-7285-6348]{Ryan M. Shannon}
\affiliation{Centre for Astrophysics and Supercomputing, Swinburne University of Technology, John St., Hawthorn, VIC 3122, Australia}
\affiliation{ARC Centre of Excellence for Gravitational Wave Discovery (OzGrav), Australia}

\author[0000-0002-1873-3718]{Fernando Camilo}
\affiliation{South African Radio Astronomy Observatory, 2 Fir Street, Observatory 7925, South Africa}

\author{John Sarkissian}
\affiliation{Australia Telescope National Facility, CSIRO, Space and Astronomy, Parkes Observatory, PO Box 276, Parkes NSW 2870, Australia}

\begin{abstract}

In addition to being the most magnetic objects in the known universe, magnetars are the only objects observed to generate fast-radio-burst-like emissions. The formation mechanism of magnetars is still highly debated, and may potentially be probed with the magnetar velocity distribution. 
We carried out a 3-year-long astrometric campaign on \swift\ --- the fastest-spinning magnetar, using the Very Long Baseline Array. After applying the phase-calibrating 1D interpolation strategy, we obtained a small proper motion of 8.5\,\maspy\ magnitude, and a parallax of $0.12\pm0.02$\,mas (uncertainties at $1\,\sigma$ confidence throughout the Letter) for \swift. The latter is the second magnetar parallax, and is among the smallest neutron star parallaxes ever determined. From the parallax, we derived the distance $9.4^{+2.0}_{-1.6}$\,kpc, which locates \swift\ at the far side of the Galactic central region. Combined with the distance, the small proper motion leads to a transverse peculiar velocity $v_\perp=48^{+50}_{-16}$\,\kmps\ --- a new lower limit to magnetar $v_\perp$. Incorporating previous $v_\perp$ estimates of seven other magnetars, we 
acquired $v_\perp=149^{+132}_{-68}$\,\kmps\ for the sample of astrometrically studied magnetars, corresponding to the three-dimensional space velocity $\sim190^{+168}_{-87}$\,\kmps, smaller than the average level of young pulsars. Additionally, we found that the magnetar velocity sample does not follow the unimodal young pulsar velocity distribution reported by Hobbs et~al. at $>2\,\sigma$ confidence, while loosely agreeing with more recent bimodal young pulsar velocity distributions derived from relatively small samples of quality astrometric determinations.

\end{abstract}

\keywords{Very long baseline interferometry (1769) --- Magnetars (992) --- Radio pulsars (1353) --- Proper motions (1295) --- Annual parallax (42)}

\section{Introduction}
\label{sec:intro}

Magnetars are a class of highly magnetized, slowly rotating neutron stars with surface magnetic field strengths typically inferred in the range $10^{14}$ -- $10^{15}$\,G, making them the most magnetic objects in the known universe. They have been observed to emit an enormous amount of high-energy electromagnetic radiation, and to undergo powerful X-ray and gamma-ray outbursts. The high energy emission from these objects is thought to be powered by the decay of their powerful magnetic fields \citep{Thompson95,Heyl98}.
A radio burst with a luminosity approaching those of fast radio bursts (FRBs) was observed from the Galactic magnetar SGR~1935$+$2154 \citep{Andersen20,Bochenek20}, which strongly reinforced the long-held speculation that a fraction (if not all) of FRBs originate from magnetars.

At the basis, the formation mechanism of magnetars is still under debate. A few theories have been proposed to explain the origin of magnetars, including core-collapse supernovae (CCSNe) \citep{Schneider19,Shenar23}, accretion-induced collapse (AIC) of white dwarfs \citep{Lipunov85,Fryer99,Dessart07,Margalit19,Ruiter19} and double neutron star (DNS) mergers \citep{Giacomazzo13,Margalit19}. 
The light curves of a few X-ray transients are believed to be generated by extragalactic millisecond magnetars born from neutron star mergers \citep[e.g.][]{Xue19,Sun19,Ai21}. 
Despite these evidences in favour of magnetar production in mergers, the DNS-merger scenario is disfavored for the bulk of the known Galactic magnetar population by their low Galactic latitudes \citep{Olausen14}. On the other hand, at least a fraction of Galactic magnetars are expected to come from CCSNe, given the 15 confirmed or proposed associations (see Table~1 of \citealp{Sherman24}) between Galactic magnetars and supernova remnants (SNRs) \citep[e.g.][]{Vasisht97,Klose04,Gelfand07,Gaensler08,Gaensler14,Borkowski17,Bailes21}. 
For formation channels other than the CCSNe or the DNS-merger channel, it remains unclear whether they would contribute to the birth of Galactic magnetars.
As the magnetar space (or peculiar) velocity (i.e., velocity with respect to its neighbourhood in the Galaxy) distribution probably varies with the underlying formation channel, it can be used to probe the formation mechanism of Galactic magnetars \citep{Ding22,Ding23a}, with an increasing number of astrometrically constrained magnetars.
In addition, pinpointing the 3D locations of magnetars in our Galaxy can help develop a template for magnetar distribution in spiral galaxies; this template can be compared against FRBs localized to spiral galaxy hosts \citep{Mannings21}, thus testing the link between FRBs and magnetars \citep{Ding22}.

To date, approximately 24 magnetars and 6 magnetar candidates have been discovered\footnote{\label{footnote:magnetar_catalog}As counted by \url{http://www.physics.mcgill.ca/~pulsar/magnetar/main.html}. The count does not include PSR~J1846$-$0258.}; however, only 6 (including \swift) have been found to be radio bright. Observations of magnetar radio pulses reveal they are quite distinct from the radio emission seen in pulsars -- they have largely flat radio spectra and their pulse profiles are highly variable on timescales ranging between seconds to years \citep[e.g.][]{Camilo08,Lower20a}. 
With a current sample size of only eight magnetars with proper motions well measured at radio \citep{Deller12a,Bower15,Ding20c} or infrared/optical wavelengths \citep{Tendulkar12,Tendulkar13,Lyman22}, the appearance of a new radio-emitting magnetar offers a valuable opportunity for study, particularly for pulsar timing and astrometry, both of which can be performed much more precisely with radio observations than with optical/infrared or X-rays \citep[e.g.][]{Kaplan08}.
Moreover, the high spatial resolution promised by radio interferometry would reduce the sample bias that favors magnetars with larger proper motions. 
It is also noteworthy that timing observations of a magnetar can, in principle, acquire position and proper motion measurements as well. However, such measurements are hampered by the extreme torque variations and associated spin-down noise of magnetars \citep[e.g.][]{Camilo07}. Hence, adopting the accurate interferometry-based position, proper motion and parallax would significantly improve the reliability and usefulness of timing observations of radio magnetars.

\subsection{\swift}
\swift\ was detected by Swift/BAT (Burst Alert Telescope) and reported on 12 March 2020 as a new soft gamma-ray repeater (SGR) and a magnetar candidate\footnote{\url{https://gcn.gsfc.nasa.gov/gcn3/27373.gcn3}}. Its identity as a magnetar was soon confirmed by the NICER team with follow-up observations \citep{Enoto20}. Subsequently, pulsed radio emission at $\approx0.7$\,mJy was detected at L band with the Effelsberg and Lovell telescopes respectively \citep{Champion20a}, showing a high dispersion measure (DM) of 706\,\pcpcc. The distance to \swift\ based on the DM is estimated to be 4.8 to 8.1\,kpc \citep{Lower20a}, according to the YMW16 \citep{Yao17} and the NE2001 \citep{Cordes02} models of the Galactic free electron distribution.
The radio emission of \swift\ was initially found to be steep-spectrum (spectral index $\alpha\lesssim-1.8$, e.g. \citealp{Lower20a, Champion20a}), which then flattened sometime in July 2020 at cm- to sub-mm-wavelengths \citep[e.g.][]{Lower20c,Torne20}.
Combining the spin period and its time derivative obtained with pulsar timing, \citet{Rajwade22} derived a characteristic age of $\sim860$\,yr, potentially making \swift\ the youngest known magnetar to date. 

The discovery of this new radio-loud magnetar offers a rare chance to refine the magnetar space velocity distribution. To obtain reliable space velocity for \swift\ requires determination of both its proper motion and distance, which can potentially be obtained with very long baseline interferometry (VLBI) observations.
In this Letter, we introduce the results of our 3-year-long astrometric campaign of \swift, and discuss their implications. Throughout the Letter, uncertainties are given at 68\% confidence, unless otherwise stated.

\section{Observations}
\label{sec:obs}

\begin{figure}
    \centering
	\includegraphics[width=\columnwidth]{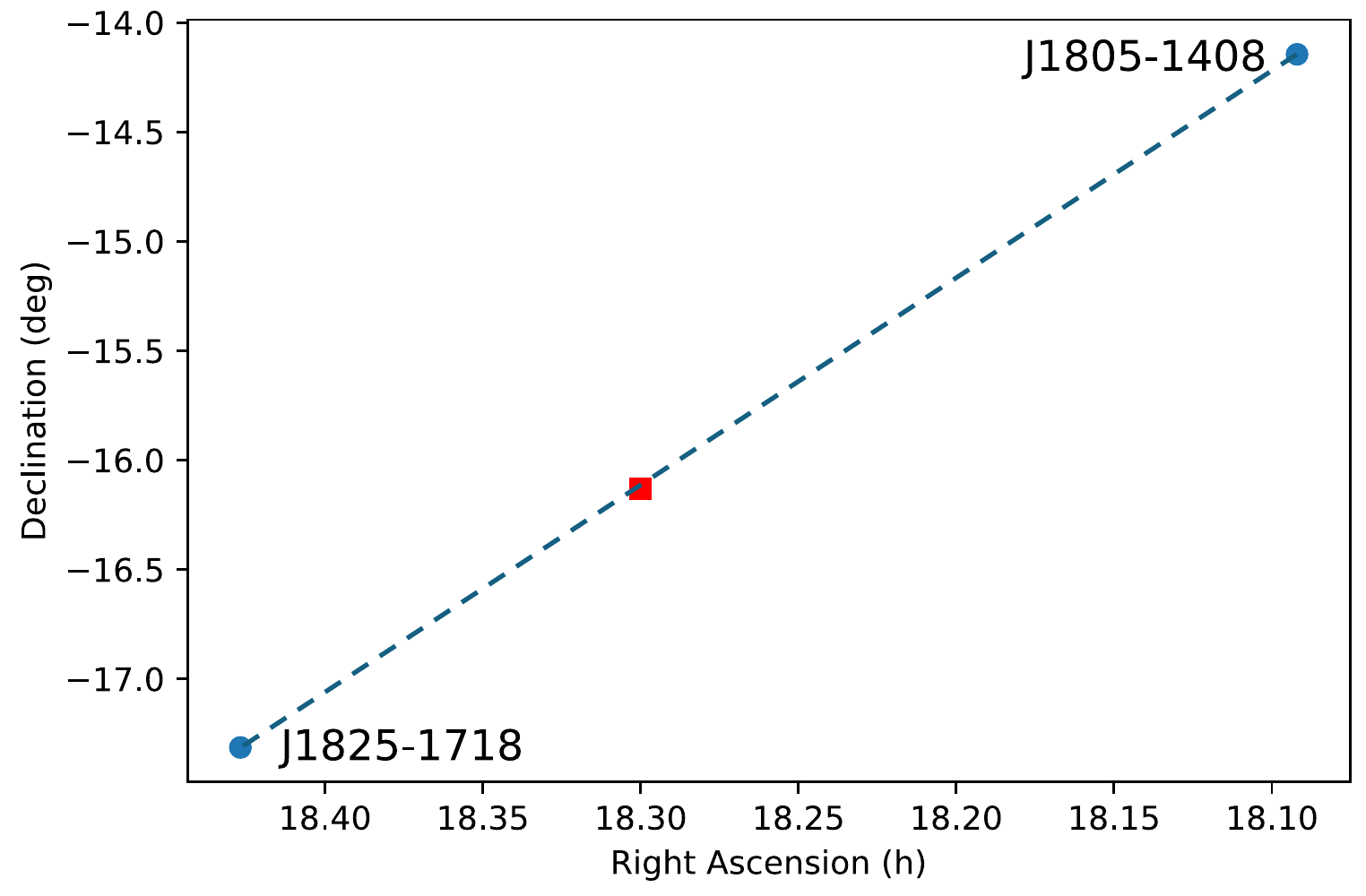}
    \caption{Calibrator configuration for the astrometry of \swift, where \swift\ is represented by the red rectangle is phase-referenced to two quasars, ICRF~J182536.5$-$171849 and ICRF~J180531.2$-$140844. The dashed line connects the two quasars, and is only $25\farcs7$ away from \swift.}
    \label{fig:calibrator_plan}
\end{figure}

Soon after the confirmation of \swift\ as a new radio magnetar \citep{Champion20a}, we proposed and acquired 3 Director's Discrepancy Time (DDT) observations from the Very Long Baseline Array (VLBA) under the project code BD232. 
To suppress the propagation-related systematic errors, the 1D interpolation strategy \citep[e.g.][]{Fomalont03,Doi06,Ding20c} was adopted in BD232 and the following observations, where \swift\ is phase-referenced to two quasars (i.e., ICRF~J182536.5$-$171849 and ICRF~J180531.2$-$140844) quasi-colinear with \swift\ (see Figure~\ref{fig:calibrator_plan}).
Given the then-steep radio spectrum \citep[e.g.][]{Lower20a,Champion20a}, the first VLBA observation (project code: BD232A) of \swift\ was made at $\sim1.6$\,GHz in April 2020. Unfortunately, no detection was achieved from the first VLBA observation (see Section~\ref{subsec:angular_broadening} for the likely reason); the $5\,\sigma$ upper limit of the BD232A observation is 0.29\,\mjypb\ \citep{Ding20b}.

After the spectral flattening of \swift\ in July 2020, we changed the observing frequency to $\sim8.8$\,GHz  to reduce the effects of angular broadening caused by ionized interstellar media (IISM), and achieved the first VLBI detection of \swift\ \citep{Ding20b}. Thereafter, we extended the astrometric campaign with two regular VLBA proposals under the project codes BD241 and BD254, and made in total 16 VLBA observations. The details of the observations are summarized in Table~\ref{tab:obs_details}.

\begin{table*}

\centering
\caption{Details of VLBA observations}
\label{tab:obs_details}

\resizebox{15cm}{!}{
\begin{tabular}{ccccccc}
\hline
\hline
Project & yyyy-mm & Central obs. freq. & Data rate & Detection? & $\mathcal{S}_\mathrm{avg}$ $^{a}$ & Gating $^{b}$ \\
code & & (GHz) & (Gbps) & & (mJy) & gain \\
\hline
BD232A & 2020-04 & 1.6 & 2 & no & --- & --- \\
BD232B & 2020-08 & 8.76 & 2 & yes & 1.3 & --- $^{c}$ \\
BD232C & 2020-11 & 8.76/15.24 $^{d}$ & 2 & yes & 1.1 & 1.4 \\
BD241A & 2021-03 & 8.76 & 2 & yes & 0.83 & 2.3  \\
BD241B & 2021-03 & 8.76 & 2 & yes & 0.48 & 2.0 \\
BD241C & 2021-09 & 8.76 & 2 & yes & 0.85 & 1.7 \\
BD241D & 2021-10 & 8.76 & 2 & yes & 1.0 & 1.9 \\
BD241E & 2022-03 & 8.76 & 2 & yes & 0.23 & 1.9 \\
BD241F & 2022-04 & 8.76 & 2 & no & --- & --- \\
BD241G & 2022-04 & 8.76 & 2 & no & --- &--- \\
BD254A & 2022-12 & 8.63 & 4 & yes & 0.36 &2.1 \\
BD254B & 2022-12 & 8.63 & 4 & yes & 0.49 & 2.0 \\
BD254C & 2023-03 & 8.63 & 4 & yes & 0.33 &4.1 \\
BD254D & 2023-03 & 8.63 & 4 & yes & 0.44 & 3.2 \\
BD254E & 2023-03 & 8.63 & 4 & yes & 0.25 & 3.6 \\
BD254F & 2023-03 & 8.63 & 4 & yes & 0.26 & 2.8 \\

\hline

\multicolumn{7}{l}{$^{a}$ $\mathcal{S}_\mathrm{avg}$ stands for the unresolved flux density averaged over the spin period.}\\
\multicolumn{7}{l}{$^{b}$ The gating gain is the ratio between the gated and the ungated image S/Ns.}\\
\multicolumn{7}{l}{$^{c}$ The pulsar gating at the BD232B epoch was unsuccessful.}\\
\multicolumn{7}{l}{$^{d}$ A dual-frequency observation was made to identify the best observing frequency. We}\\
\multicolumn{7}{l}{\,\,\,\,eventually gave up observing at 15\,GHz due to the absence of suitable calibrator plan.}\\

\end{tabular}

}

\end{table*}

Our VLBA campaign of \swift\ was supported by high-cadence pulsar timing observations of \swift\ using the Murriyang/Parkes 64-meter radio telescope: ongoing changes in the flux density and pulse ephemeris of \swift\ were provided from the timing observations. 
Generally, \swift\ is a faint target with $\lesssim1$\,mJy unresolved flux density averaged over the spin period (see Table~\ref{tab:obs_details}). 
Based on the pulse ephemerides, pulsar gating was implemented at correlation (of the VLBA data) using the {\tt DiFX} software correlator \citep{Deller11a}, which typically improves the image S/N of \swift\ by a factor of $\sim2$ (see Table~\ref{tab:obs_details}).

In addition, the flux density of \swift\ is highly variable. For instance, the flux density dropped below the detection limit of VLBA between April 2022 and November 2022. 
In response to the decline of flux density, the data recording rate of the BD254 observations was increased from 2\,Gbps to the highest 4\,Gbps. As a result, the central observing frequency changes  by 0.1\,GHz (see Table~\ref{tab:obs_details}). The resultant position shift of \swift\ due to the frequency-dependent core shifts (FDCSs; e.g. \citealp{Sokolovsky11}) of the phase calibrators is at the order of $\lesssim2\,\mu$as. Therefore, the slightly different central observing frequency in BD254 observations has negligible impact on the astrometric results.
In December 2022, the flux density of \swift\ rebounded, which, however, only lasted for about 3 months. Since early 2023, the flux density of \swift\ had been declining again (Lower et~al. in prep.), then faded below the sensitivity of the VLBA. 
Consequently, the VLBA observations originally planned for September 2023 were canceled; the final VLBA observations reported in the Letter were made in March 2023.

\section{Data reduction \& Direct results}
\label{sec:data_reduction}

We used the {\tt psrvlbireduce}\footnote{\url{https://github.com/dingswin/psrvlbireduce}} pipeline for the data reduction of the VLBA data. The pipeline is written in {\tt ParselTongue}, an interface connecting {\tt AIPS} \citep{Greisen03} and {\tt python}. 
As mentioned in Section~\ref{sec:obs}, the 1D interpolation strategy was adopted in the astrometric campaign of \swift.
Prior to this astrometric campaign, the same strategy has been used in the VLBI astrometry of the radio-loud magnetar XTE~J1810$-$197, which led to the first magnetar parallax \citep{Ding20c} (D20). 
The reduction of the \swift\ data follows the same procedure as D20: 
the wrap of the residual phase between the two phase calibrators was solved in an iterative style; the set of corrections leading to the largest image S/N of the magnetar is adopted as the solution.
Compared to D20, acquiring the phase-wrap solutions is straightforward in this campaign, thanks to the higher observing frequency and the smaller angular distance ($5\fdg8$) between the two phase calibrators. 
Observing at higher frequency significantly reduces the impact of IISM-induced angular broadening (see Section~\ref{subsec:angular_broadening}).
As a result, much better phase solutions were obtained at longer baselines, and no data from any station were excluded from the subsequent analysis.
The image models of the two phase calibrators are provided online\footnote{available on Zenodo under an open-source 
Creative Commons Attribution license:
\dataset[doi:10.5281/zenodo.11239303]{https://doi.org/10.5281/zenodo.11239303}}, in order to convenience the reproduction of our results.

After the data reduction, we detected \swift\ in 13 of the 16 VLBA observations (see Table~\ref{tab:obs_details}). From the final image of \swift\ acquired at each epoch of detection, we obtained one $\sim\!8.7$-GHz position and its statistical uncertainty $\sigma^\mathcal{R}_{ij}$ (where $i=\alpha, \delta$ denotes right ascension and declination, $j=1, 2, 3,...$ indicates different epochs) of \swift\ using {\tt JMFIT} (of the {\tt AIPS} package), which is compiled in Table~\ref{tab:position_series}.

\begin{table}
\centering
\caption{Position series of \swift}
\label{tab:position_series}
\begin{tabular}{ccc}
\hline
\hline
 Epoch & Right ascension & Declination  \\
 (yr) &  &  \\
\hline
2020.6314 & $18^{\rm h}18^{\rm m}00\fs193404(2|5)$ & $-16\degr07'53\farcs00499(5|15)$ \\
2020.8739 & $18^{\rm h}18^{\rm m}00\fs193349(3|5)$ & $-16\degr07'53\farcs00703(9|23)$ \\
2021.1906 & $18^{\rm h}18^{\rm m}00\fs193285(2|5)$ & $-16\degr07'53\farcs00930(6|22)$ \\
2021.2371 & $18^{\rm h}18^{\rm m}00\fs193269(3|5)$ & $-16\degr07'53\farcs00956(8|14)$ \\
2021.6933 & $18^{\rm h}18^{\rm m}00\fs193140(2|4)$ & $-16\degr07'53\farcs01286(6|14)$ \\
2021.7780 & $18^{\rm h}18^{\rm m}00\fs193126(2|4)$ & $-16\degr07'53\farcs01394(5|12)$ \\
2022.2179 & $18^{\rm h}18^{\rm m}00\fs193029(6|7)$ & $-16\degr07'53\farcs0174(2|2)$ \\
2022.9393 & $18^{\rm h}18^{\rm m}00\fs192837(3|4)$ & $-16\degr07'53\farcs02306(8|15)$ \\
2022.9420 & $18^{\rm h}18^{\rm m}00\fs192837(3|5)$ & $-16\degr07'53\farcs0227(1|3)$ \\
2023.1633 & $18^{\rm h}18^{\rm m}00\fs192795(2|5)$ & $-16\degr07'53\farcs02436(7|22)$ \\
2023.2015 & $18^{\rm h}18^{\rm m}00\fs192778(3|5)$ & $-16\degr07'53\farcs02470(8|33)$ \\
2023.2261 & $18^{\rm h}18^{\rm m}00\fs192786(3|5)$ & $-16\degr07'53\farcs02475(8|17)$ \\
2023.2425 & $18^{\rm h}18^{\rm m}00\fs192775(2|4)$ & $-16\degr07'53\farcs02516(5|12)$ \\

\hline

\end{tabular}
\tablenotetext{}{\raggedright The errors on the left and right side of ``$|$'' represent, respectively, statistical uncertainty (from the image-plane position fit) and the total uncertainty adding a fiducial systematic component to the statistical component (see Section~\ref{sec:astrometric_inference}).}

\end{table}

\subsection{Angular broadening of \swift}
\label{subsec:angular_broadening}

From the final images of \swift, the apparent size of \swift\ can also be constrained. Pulsars are point-like sources, so their apparent sizes are normally consistent with zero \citep[e.g.][]{Cordes83}. The VLBI image of a pulsar only becomes resolved when multi-path propagation of its radio emissions occurs due to scattering by IISM \citep[e.g.][]{Bower14,Ding23}. 
By the method described in Appendix~A of D20, we estimated a half width $\theta_\mathrm{sc}=1.08\pm0.37$\,mas of the angular-broadened size for \swift\ at $\sim8.7$\,GHz.
As $\theta_\mathrm{sc}$ theoretically changes with the observing frequency $\nu$ as  $\nu^{-11/5}$ assuming a thin-screen distribution of the foreground IISM \citep{Goodman89,Macquart13}, the $\theta_\mathrm{sc}$ at 1.6\,GHz is $\sim41.5\times\left(1.08\pm0.37\right)\,\mathrm{mas}=45\pm15$\,mas. Therefore, the non-detection of the BD232A observation was likely mainly caused by the severe angular broadening at $\sim1.6$\,GHz.

\section{Systematic errors \& Astrometric inference}
\label{sec:astrometric_inference}

The astrometric parameters, including the reference position, proper motion, and parallax, can be inferred from the position series (see Table~\ref{tab:position_series}) of \swift.
In addition to the statistical positional uncertainties $\sigma^\mathcal{R}$ (i.e., the uncertainties due to random thermal noise in the image) that can be evaluated from the final \swift\ images, systematic errors mainly caused by atmospheric propagation effects also contribute to the error budget of the \swift\ positions provided in Table~\ref{tab:position_series}.
In three different approaches, we carried out astrometric inference that accounts for the presence of systematic errors.

In the first approach, we derived astrometric parameters along with the fiducial systematic uncertainties $\sigma^\mathcal{S}$ using the same least-squares method described in Section~4.2 of D20, except that the coefficient $A$ in Equation~2 of D20 was directly estimated with respect to the virtual calibrator.
We obtained $A=6.79\times10^{-2}$ for this work. 
The corresponding total uncertainties calculated as $\sqrt{\left(\sigma^\mathrm{R}\right)^2+\left(\sigma^\mathrm{S}\right)^2}$ are dominated by $\sigma^\mathcal{S}$, and provided on the right side of the `` $|$ '' marks in Table~\ref{tab:position_series}. 
The resultant astrometric fit is given in Table~\ref{tab:mu_and_pi}.
In the second approach, we performed bootstrap \citep{Efron94} analysis as described in Section~3.1 of D20, based on the error recipe obtained with the first approach. To be independent of the first approach, we also carried out bootstrapping based on only  $\sigma^\mathcal{R}$. 
The results of the two bootstrap realizations can be found in Table~\ref{tab:mu_and_pi}.

{\renewcommand{\arraystretch}{1.5}
\begin{table*}
	\centering
	\caption{Proper motion and parallax measurements for \swift}
	\label{tab:mu_and_pi}
	\begin{tabular}{ccccccc} 
		\hline
            \hline
	method & $\sigma_{ij}$  & $\eta_\mathrm{EFAC}$ & $\eta_\delta$ & $\mu_\alpha \equiv \dot{\alpha}\cos{\delta}$ & $\mu_\delta$ & $\varpi$   \\
	 &  & & & (\maspy) & (\maspy) & (mas)   \\
		\hline
	least-square fitting & $\sqrt{\left(\sigma^\mathcal{R}_{ij}\right)^2+\left(\sigma^\mathcal{S}_{ij}\right)^2}$ & --- & --- & $-3.570\pm0.016$ & $-7.723\pm0.035$ & $0.121\pm0.018$  \\
	bootstrap & $\sigma^\mathcal{R}_{ij}$ & --- & --- & $-3.575^{\,+0.018}_{\,-0.015}$ & $-7.694^{\,+0.039}_{\,-0.038}$ & $0.122^{\,+0.026}_{\,-0.023}$  \\
    bootstrap & $\sqrt{\left(\sigma^\mathcal{R}_{ij}\right)^2+\left(\sigma^\mathcal{S}_{ij}\right)^2}$ & --- & --- & $-3.572\pm0.015$ & $-7.720^{\,+0.051}_{\,-0.050}$ & $0.123^{\,+0.022}_{\,-0.020}$  \\
    Bayesian & $\sqrt{\left(\sigma^\mathcal{R}_{ij}\right)^2+\left(\eta_\mathrm{EFAC} \cdot \sigma^\mathcal{S}_{ij}\right)^2}$ & $1.04^{\,+0.25}_{\,-0.20}$ & --- & $-3.571^{\,+0.024}_{\,-0.023}$ & $-7.723\pm0.053$ & $0.122\pm0.026$  \\
    Bayesian & $\sqrt{\left(\sigma^\mathcal{R}_{ij}\right)^2+\left(\eta'_{i} \cdot \sigma^\mathcal{S}_{ij}\right)^2}$ & $0.73^{\,+0.36}_{\,-0.31}$ & $1.7^{\,+1.5}_{\,-0.7}$ & $-3.572\pm0.019$ & $-7.724^{\,+0.060}_{\,-0.063}$ & $0.121^{\,+0.020}_{\,-0.021}$  \\
		\hline
	\end{tabular}
    \tablenotetext{}{\raggedright $\sigma_{ij}$ stands for the total positional uncertainties, where $i=\alpha, \delta$ refers to right ascension or declination, and $j=1, 2, 3,...$ specifies an observation. $\eta_\mathrm{EFAC}$ and $\eta$ are the systematics-correcting factors defined in Section~\ref{subsec:two_nuisance_parameters}. $\mu_\alpha$ and $\mu_\delta$ represent, respectively, the right ascension and the declination component of the proper motion. $\varpi$ denotes the parallax. The estimates at the bottom are adopted as the final results (see Section~\ref{subsec:two_nuisance_parameters} for justifications).}
\end{table*}
}

In the last approach, we made Bayesian astrometric inference using {\tt sterne.py}\footnote{The {\tt sterne.py} version utilized in this work is made publicly available at \url{https://doi.org/10.5281/zenodo.11239560}.} \citep{Ding24a}. Following \citet{Ding23}, we estimated the systematic errors by introducing the correction factor $\eta_\mathrm{EFAC}$ as
\begin{equation}
\label{eq:EFAC}
    \sigma_{ij}\left(\eta_\mathrm{EFAC}\right)=\sqrt{(\sigma_{ij}^\mathcal{R})^2+(\eta_\mathrm{EFAC} \cdot \sigma_{ij}^\mathcal{S})^2} \,,
\end{equation}
where $i=\alpha, \delta$ denotes right ascension (RA) or declination, $j=1, 2, 3,...$ indicates different epochs.
The results of the Bayesian inference are provided in Table~\ref{tab:mu_and_pi}. 
Furthermore, in this work, we introduce an extra nuisance parameter into the Bayesian inference in order to better characterize the systematic errors, which is explained as follows.

\subsection{Inference of systematics: 2 better than 1}
\label{subsec:two_nuisance_parameters}
\begin{figure*}
    \centering
    \includegraphics[width=0.75\textwidth]{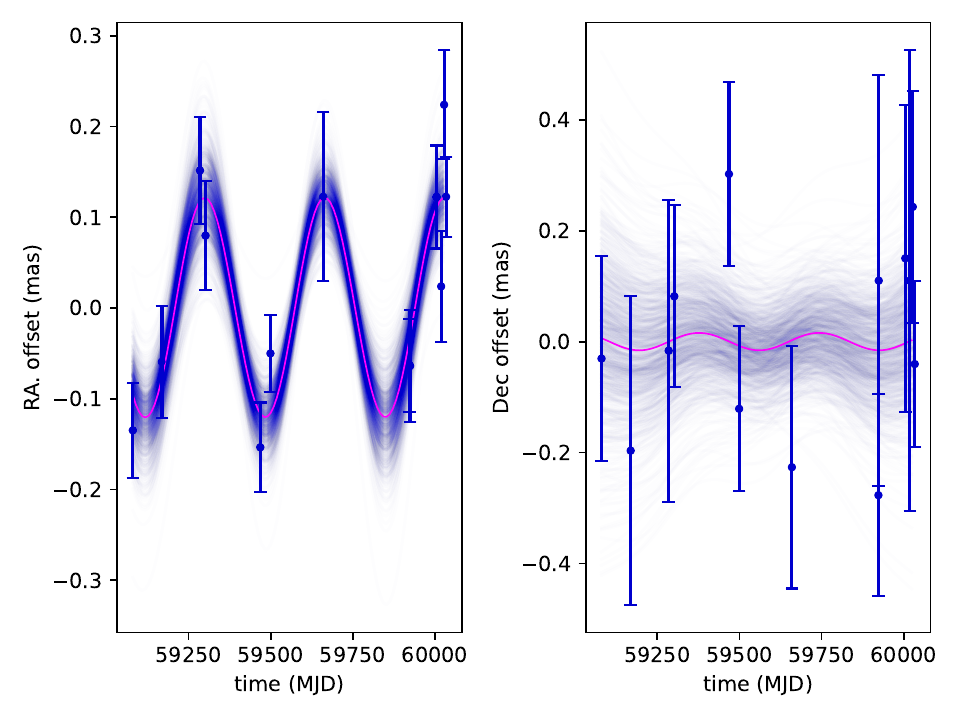}
    \caption{Position evolution of \swift\ with the best-fit proper motion effects removed. The positional uncertainties are calculated with Equation~\ref{eq:two_nuisance_parameters}, and already include the effects of the two nuisance parameters correcting the systematic errors. The inferred astrometric model is shown with the pink curve, while the overlaid Bayesian simulations visualize the uncertainty of the model.}
    \label{fig:parallax_signature}
\end{figure*}

The estimation of $\sigma_{ij}^\mathcal{S}$ based on Equation~2 of D20 assumes $\sigma_{\alpha j}^\mathcal{S}/\sigma_{\delta j}^\mathcal{S}=\Theta_{\alpha j}/\Theta_{\delta j}$, where $\Theta_{ij}$ refers to the synthesized beam size projected to the $i$-th direction.
This assumption is not necessarily true. In particular, in low-elevation observations (which is the case for \swift\ observations with VLBA), any change in the declination of pointing (during the source switches) would cause disproportionately larger path length difference (thus leading to disproportionately larger propagation-related systematics), as compared to any change in the RA of pointing. 
Therefore, we introduced an extra nuisance parameter $\eta_\delta$ (i.e., {\tt EFAD} in {\tt sterne.py}), and infer the systematics as
\begin{equation}
\label{eq:two_nuisance_parameters}
    \sigma_{ij}\left(\eta'_\mathrm{i}\right)=\sqrt{(\sigma_{ij}^\mathcal{R})^2+(\eta'_\mathrm{i} \cdot \sigma_{ij}^\mathcal{S})^2} \,,
\end{equation}
where $\eta'_\alpha=\eta_\mathrm{EFAC}$, and $\eta'_\delta=\eta_\mathrm{EFAC}\cdot\eta_\delta$. When the inference of $\eta_\delta$ is not requested, Equation~\ref{eq:two_nuisance_parameters} returns to Equation~\ref{eq:EFAC}.
The results of the Bayesian inference that includes both $\eta_\mathrm{EFAC}$ and $\eta_\delta$ are listed in Table~\ref{tab:mu_and_pi}. The parallax signature revealed with the two-nuisance-parameter (hereafter abbreviated as 2NP) Bayesian inference is illustrated in Figure~\ref{fig:parallax_signature}, where the positional uncertainties are already updated according to Equation~\ref{eq:two_nuisance_parameters}.

As described in this section, we made astrometric inference that factors in systematic errors by different methods and error recipe. As shown in Table~\ref{tab:mu_and_pi}, all the methods and the error recipes offer consistent astrometric results.
Following the discussion in Section~4.2 of \citet{Ding23} based on an astrometric sample of 18, we continue to adopt Bayesian results in this work. 

Between the two Bayesian realizations, we expect the 2NP inference to more accurately characterize the systematics (as compared to using one nuisance parameter), at an acceptable cost (5\%) of statistical significance. 
Therefore, we use the results of the 2NP inference for the discussions that follow in this Letter.
As shown in Table~\ref{tab:mu_and_pi}, the 2NP Bayesian inference suggests that the one-nuisance-parameter (hereafter abbreviated as 1NP) inference overstates $\sigma_{\alpha j}^\mathcal{S}$ by a factor of $\sim1.4$, while underestimating $\sigma_{\delta j}^\mathcal{S}$ by a factor of $\sim1.2$.
Accordingly, the proper motion in RA $\mu_\alpha$ is more precise with the 2NP inference, while the uncertainty on the proper motion in declination $\mu_\delta$ gets larger. 
Because the parallax effect is much more prominent in RA (than in declination), the reduction in uncertainties of the RA measurements in this approach means that higher parallax precision results from the 2NP Bayesian inference, as compared to the 1NP inference.

\subsection{The absolute position of \swift}
\label{subsec:absolute_position}

The full astrometric model must also include the reference position of \swift. Due to the relative astrometry nature of this work, the reference position provided by any astrometric inference can only be considered as a relative position with respect to the phase calibrator(s).
To derive the absolute position (of \swift) that can be compared with a position measured elsewhere (e.g., with pulsar timing), we followed the procedure described in Section~4.4 of \citet{Ding20c}, and obtained the absolute position $18^\mathrm{h}18^\mathrm{m}00\fs193170(9), -16\degr07'53\farcs0190(2)$ at the reference epoch of MJD~59660. 
Here, the positional uncertainty includes {\bf 1)} the positional uncertainties of the two reference sources, {\bf 2)} the residual first-order propagation-related systematic uncertainty, and {\bf 3)} the position offset induced by the FDCSs of the reference sources.
The absolute position is based on the calibrator positions $18^\mathrm{h}25^\mathrm{m}36\fs532303\pm0.12\,\mathrm{mas}, -17\degr18'49\farcs84746\pm0.17\,\mathrm{mas}$ and $18^\mathrm{h}05^\mathrm{m}31\fs23753\pm0.26\,\mathrm{mas}, -14\degr08'44\farcs68657\pm0.44\,\mathrm{mas}$ reported for ICRF~J182536.5$-$171849 and ICRF~J180531.2$-$140844, respectively, in the latest 2024a release of the Radio Fundamental Catalogue\footnote{\url{https://astrogeo.org/}}. 
Thanks to the relatively high observing frequency ($\sim8.7$\,GHz) of this work, the position offset due to FDCSs is $\lesssim0.06$\,mas \citep{Sokolovsky11}, thus $\lesssim0.04$\,mas in either right ascension or declination, much smaller than the $\sim0.8$\,mas level at $\sim1.5$\,GHz \citep{Sokolovsky11,Ding20} where most astrometric campaigns of pulsars are carried out \citep[e.g.][]{Deller19,Ding23}.

\section{Distance \& velocity}
\label{sec:D_and_v_t}

The determination of both parallax and proper motion of \swift\ enables us to derive its distance and the space velocity, which are desired for achieving the scientific goals outlined in Section~\ref{sec:intro}. 
The space velocity refers to the relative velocity with respect to the stellar neighbourhood, which can be linked to the natal kick received at the NS birth. As no information of the radial velocity $v_\mathrm{r}$ is available, we can only constrain the space velocity tangential to the line of sight, hereafter referred to as the transverse space velocity $v_\perp$.

\subsection{The distance to \swift}
\label{subsec:distance}
\begin{figure}
    \centering
    \includegraphics[width=\columnwidth]{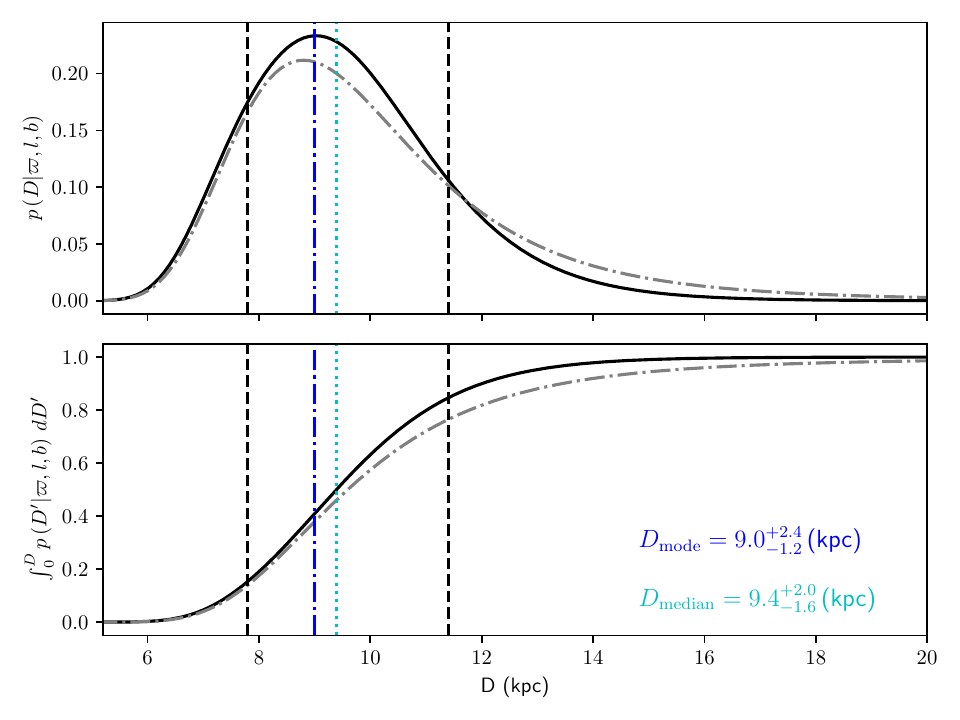}
    \caption{The distance $D$ of \swift\ inferred from the parallax $\varpi$, the Galactic longitude $l$ and Galactic latitude $b$ (see Section~\ref{sec:D_and_v_t}). The solid curves in the upper and lower panel represent, respectively, the probability density function (PDF) and the cumulative distribution function (CDF) of distance. The dashed vertical lines correspond to 0.16 and 0.84 of the CDF, and enclose the $1\,\sigma$ uncertainty interval of the distance. For comparison, the PDF and CDF disregarding prior spatial distribution are shown with the dash-dotted grey curves.}
    \label{fig:distance}
\end{figure}

To infer a parallax-based distance normally requires a prior Galactic spatial distribution \citep[e.g.][]{Igoshev16,Bailer-Jones21}, although the impact of the prior choice approaches negligible levels once the parallax is measured with $\gtrsim7\,\sigma$ significance \citep{Lutz73}. 
With no Galactic spatial distribution available for magnetars, we instead use the Galactic spatial distribution of pulsars (``Model C'') provided by \citet{Lorimer06a}.
Following the procedure described in Section~6.1 of \citet{Ding23}, we obtained the distance $9.4^{+2.0}_{-1.6}$\,kpc to \swift.
Unlike the millisecond pulsars discussed in \citet{Ding23}, Galactic magnetars are much younger objects situated close to the Galactic plane (see Figure~1 of \citealp{Ding23a}). Therefore, we adopted the original parameter ``$E$'' (i.e., 0.18\,kpc) of the ``Model C'' given by \citet{Lorimer06a} for the prior spatial distribution. 
The probability density function (PDF) and cumulative distribution function (CDF) of distance are plotted in Figure~\ref{fig:distance}.
For comparison, the PDF and CDF derived without assuming any prior spatial distribution are overlaid to Figure~\ref{fig:distance}. 
We found that the use of the pulsar spatial distribution \citep{Lorimer06a} slightly enlarges the distance estimate while reducing the skewness of the PDF. However, the impact of the prior spatial distribution (on the distance inference) is very limited thanks to the relatively high parallax significance.

\subsection{The transverse space velocity of \swift}
\label{subsec:v_t_J1818}

The transverse space velocity $v_\perp$ of \swift\ was estimated by the same method detailed in Section~6.2 of \citet{Ding23}. At a Galactic latitude $b$ of only $-0\fdg14$, the motion (and hence the velocity) of the star field surrounding \swift\ can be safely approximated by circular motion on the Galactic plane around the Galactic Center. We obtained $v_\perp=48^{+50}_{-16}$\,\kmps\ for \swift.

\section{Discussions}
\label{sec:discussions}

\begin{figure*}
    \centering
    \includegraphics[width=0.8\textwidth]{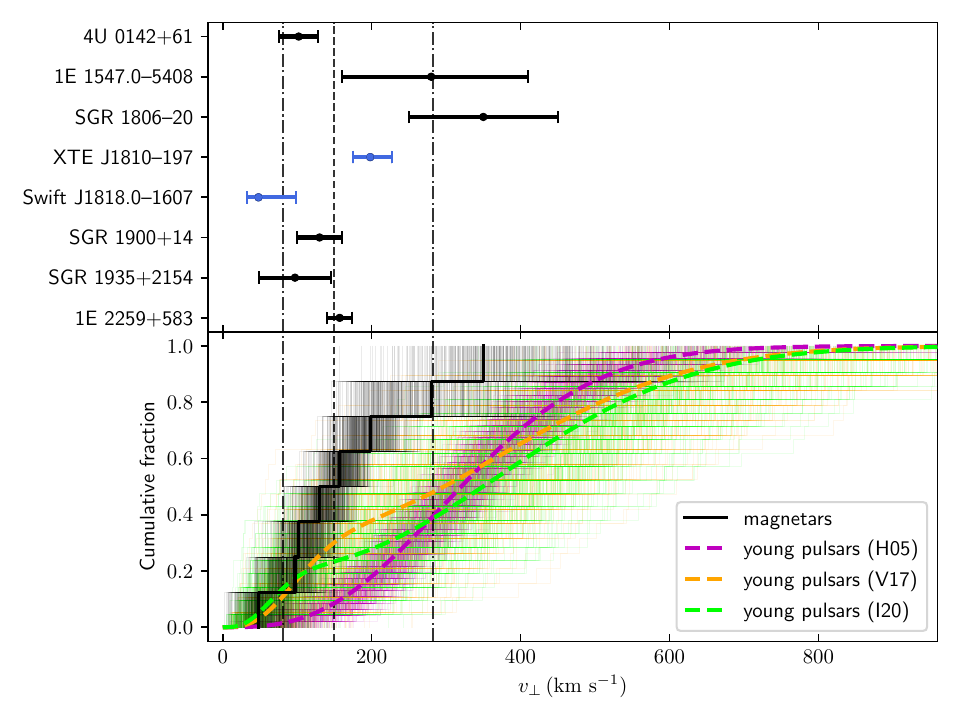}
    \caption{{\bf Upper:} The transverse space velocities $v_\perp$ of 8 magnetars, including the seven $v_\perp$ estimates historically reported by \citet{Deller12a,Tendulkar12,Tendulkar13,Ding20c,Lyman22}. Among the 8 magnetars, proper motions are constrained for all sources with either infrared/optical \citep{Tendulkar12,Tendulkar13,Lyman22} or VLBI astrometry \citep{Helfand07,Deller12a,Ding20c}, while model-independent parallax-based distances are determined for two magnetars \citep{Ding20c}, for which the $v_\perp$ estimates are highlighted in blue. 
    {\bf Lower:} The cumulative fraction of the magnetar $v_\perp$ estimates is shown with a thick black stepped lines, while thin black ones based on simulations visualize the uncertainty of the cumulative fraction. For comparison, the cumulative fractions of three $v_\perp$ distributions of young pulsars, converted from the respective 3-dimensional velocity distributions \citep{Hobbs05,Verbunt17,Igoshev20}, are overlaid with different colors. The colored dashed curves display the best-fit cumulative distribution functions (CDFs). The thin colored stepped lines show simulations based on the CDFs, while taking into account the uncertainties of CDF parameters.
    The overall $v_\perp$ magnitude is estimated with Monte Carlo simulation. The 16th, 50th and 84th percentiles of the $v_\perp$ simulations, marked with the vertical lines, give the estimate $v_\perp=149^{+132}_{-68}$\,\kmps.}
    \label{fig:v_t_distribution}
\end{figure*}

This work has led to the second annual-geometric-parallax-based distance of a magnetar, which is among the largest pulsar distances measured at high significance (see the Catalogue of Pulsar Parallaxes\footnote{\url{https://hosting.astro.cornell.edu/research/parallax/}}). 
According to the distance presented here, \swift\ is most likely situated at the far side of the Galactic central region densely populated with stars.
In addition, the $v_\perp$ of \swift\ is the smallest one ever determined for a magnetar (see Figure~\ref{fig:v_t_distribution}). These new astrometric results have several implications, which we discuss below.

\subsection{Comparison with DM-based distances}
\label{subsec:DM_distance}
The DM of \swift\ is 699.2(8)\,\pcpcc\ \citep{Oswald21}. 
Using the {\tt pygedm}\footnote{\url{https://github.com/FRBs/pygedm}} package \citep{Price21}, we obtained DM-based distances $8.0\pm2.0$\,kpc and $4.8\pm1.2$\,kpc for \swift, based on, respectively, the NE2001 \citep{Cordes02} and the YMW16 model \citep{Yao17} --- the two latest models of Galactic free-electron distribution $n_\mathrm{e}$.  
Here, indicative 25\% fraction uncertainties are prescribed to both DM-based distances.
Our new trigonometric distance $9.4^{+2.0}_{-1.6}$\,kpc agrees with and favors the NE2001 prediction, while being larger than the YMW16-model-based distance with  $\sim2\,\sigma$ confidence.

\subsection{The upper limit to the quiescent X-ray luminosity}
\label{subsec:L_qui}

So far, X-ray luminosities or burst fluences of \swift\ have been calculated with the relatively small DM-based distances, thus being systematically under-estimated.
For instance, the upper limit of the quiescent X-ray luminosity $L_\mathrm{qui}$ of \swift\ is estimated to be $5.5\times10^{33}$\,\ergps\ assuming the YMW16-model-based 4.8\,kpc distance, which is lower than expected for a young magnetar by an order of magnitude \citep{Esposito20}.
The new trigonometric distance refines the upper limit (of the $L_\mathrm{qui}$ of \swift) to $2.1\times10^{34}$\,\ergps, which is now more consistent with expectation ($\gtrsim6\times10^{34}$\,\ergps, \citealp{Vigano13,Esposito20}). Additionally, the refined $L_\mathrm{qui}$ upper limit remains lower than the spin-down luminosity of $\sim10^{36}$\,\ergps\ \citep{Esposito20}, leaving it ambiguous if \swift\ is mainly rotationally or magnetically powered (as questioned by \citealp{Rea12,Esposito20}).

\subsection{The upper limit to the distance of the first scattering IISM screen}
\label{subsec:D_sc}

In Section~\ref{subsec:angular_broadening}, we derived the half width $\theta_\mathrm{sc}=45\pm15$\,mas of the angular-broadened size (of \swift) at 1.6\,GHz induced by IISM scattering. 
In addition to the angular broadening effect, IISM may also lead to the broadening of the pulse profile of a pulsar, which can be characterized by temporal pulse broadening $\tau_\mathrm{sc}$ \citep{Cordes85,Cordes91,Cordes05,Ocker22}. 
Assuming the IISM scattering is dominated by one thin IISM screen, the distance to the dominant IISM screen can be determined geometrically. 
Incorporating $\tau_\mathrm{sc}=8.8\pm0.2$\,ms (of \swift) at 1.6\,GHz \citep{Lower20a} and the new distance to \swift, we calculate our distance to the dominant scattering IISM screen (in front of \swift) to be $2.6^{+1.8}_{-0.9}$\,kpc using Equation~16 of \citet{Ding23}.

In a more realistic model, there must be multiple scattering IISM screens. In this case, we can place the upper limit $2.6^{+1.8}_{-0.9}$\,kpc to the distance of the first (closest to us) IISM screen, according to Equation~20 of \citet{Ding23}.

\subsection{On a proposed supernova remnant association}
\label{subsec:SNR}

As mentioned in Section~\ref{sec:intro}, the formation mechanism of magnetars is still under debate. It is believed that at least some Galactic magnetars were born in CCSNe, given their associations with SNRs (see the McGill Online Magnetar Catalog\textsuperscript{\ref{footnote:magnetar_catalog}}, \citealp{Olausen14}).
Therefore, searching for associated SNRs is essential for investigating magnetar formation channels.
Additionally, CCSN association is important for understanding magnetar evolution, and whether they are evolutionarily connected to other NS species (e.g. XDINS, see \citealp{Keane08,Benli16}).

The low proper motion and the extraordinary youth of \swift\ suggest that a putative SNR associated with \swift, if existent, must be centered around \swift. The recently discovered semi-circular radio feature around \swift\ \citep{Ibrahim23} serves as a prime candidate of the putative SNR associated with \swift. 
According to \citet{Ibrahim23}, if the semi-circular radio feature is indeed a part of an SNR, it is expected to be $\gtrsim8$\,kpc away (from us) to have entered the full Sedov phase \citep{Truelove99,Sedov18}, which is consistent with our new magnetar distance of around 9\,kpc. Therefore, the new magnetar distance sustains the interpretation that the semi-circular radio feature is the SNR associated with \swift.

\subsection{Revisiting the magnetar space velocity distribution}
\label{subsec:v_t_distribution}

Previously, 8 magnetars have been astrometrically studied at X-rays \citep{Kaplan08}, infrared/optical, \citep{Tendulkar12,Tendulkar13,Lyman22} or radio \citep{Helfand07,Deller12a,Bower15,Ding20c}, which achieved proper motion constraints for the 8 magnetars, and determined 1 significant magnetar parallax \citep{Ding20c}.
To focus on the $v_\perp$ distribution of field (or isolated) magnetars, we remove SGR~J1745$-$2900 believed to be associated with Sgr~A$^*$ \citep[e.g.][]{Bower15} from the list of the magnetars.
The $v_\perp$ estimates reported for the remaining 7 magnetars are compiled in Figure~\ref{fig:v_t_distribution}. 

This work obtains the second highly desired significant magnetar parallax, and leads to a robust $v_\perp$ determination.
With hitherto the smallest magnetar $v_\perp$ added to the existing sample of 7, we revisit the magnetar $v_\perp$ distribution.
Given that a few $v_\perp$ uncertainties are asymmetric, we estimate the overall magnetar $v_\perp$ magnitude with Monte-Carlo simulations described in Section~6.2 of \citet{Ding23}.
From the $v_\perp$ simulations, $v_\perp=149^{+132}_{-68}$\,\kmps\ is estimated for the whole sample of astrometrically studied magnetars, where the median of the $v_\perp$ simulations is adopted as the overall $v_\perp$ estimate, and the $v_\perp$ values at the 16th and the 84th percentiles of the $v_\perp$ simulations define the $1\,\sigma$ uncertainty interval.
Assuming that the directions of the magnetar 3D space velocities $v$ are uniformly random, it is easy to calculate that $v\sim4/\pi \cdot\ v_\perp=190^{+168}_{-87}$\,\kmps, which is smaller than the mean 3D velocity of young ($\tau_\mathrm{c}<3$\,Myr) pulsars ($400\pm40$\,\kmps) estimated by \citet{Hobbs05}.

Furthermore, we compare the cumulative fraction of magnetar $v_\perp$ to three empirical velocity distributions of young pulsars. The three reference velocity distributions are {\bf 1)} the Maxwellian distribution with the scale parameter $\sigma^\mathrm{Max}=265$\,\kmps\ derived by \citet{Hobbs05} (H05), {\bf 2)} the bimodal distribution reported in Table~4 (``Isotropic models'', ``Y'') of \citet{Verbunt17} (V17), and {\bf 3)} the bimodal distribution offered at the bottom of Table~2 of \citet{Igoshev20} (I20). All the three reference distributions are provided for 3-dimensional velocities $v$, which are converted to $v_\perp$ distributions using the aforementioned relation $v_\perp \sim \pi/4 \cdot v$.

In practice, we examine the null hypothesis that the magnetar $v_\perp$ sample follows a given young pulsar $v_\perp$ distribution, using the Kolmogorov–Smirnov (KS) test and the Anderson–Darling (AD) test, one after another. 
Specifically, we simulate 8 magnetar $v_\perp$ assuming normal distribution (or split normal distribution when the uncertainty is asymmetric). 
On the opposite side, $N_\mathrm{s}$ simulations are drawn from a reference distribution, where $N_\mathrm{s}$ is the size of the sample from which the original reference distribution is derived (see Table~\ref{tab:ks_and_ad_tests} for $N_\mathrm{s}$).
We make KS/AD test with the two simulated $v_\perp$ series, and record the p-value $\zeta$ of the null hypothesis. This procedure is repeated 10,000 times, which leads to 10,000 $\zeta$.
The results of the KS/AD tests are provided in Table~\ref{tab:ks_and_ad_tests}. 
We find significant evidence suggesting that the magnetar $v_\perp$ sample does not follow the unimodal young pulsar velocity distribution proposed by H05, with at least 95\% of the simulations rendering $\zeta<0.05$.
On the other hand, there is no sufficient evidence for ruling out the possibility that the magnetar $v_\perp$ sample follows either bimodal young pulsar velocity distribution that is derived from a much smaller but more accurate velocity sample compared to H05.
The cumulative fractions of the magnetar $v_\perp$ sample and the aforementioned young pulsar $v_\perp$ distribution are illustrated in Figure~\ref{fig:v_t_distribution}, overlaid with the related simulations.

\begin{table}
\centering
\caption{Results of Kolmogorov–Smirnov and Anderson–Darling tests of the magnetar $v_\perp$ sample against velocity distribution of young pulsars}
\label{tab:ks_and_ad_tests}
\begin{tabular}{c|c|c|c|c|c}
\hline
\hline
 ref.\tablenotemark{a}  & $N_\mathrm{s}$ & \multicolumn{2}{c|}{KS test}  & \multicolumn{2}{c}{AD test} \\
 \cline{3-6} 
 distr.  &  & $\zeta$ \tablenotemark{b} & $f_{\zeta<0.05}$ & $\zeta$ & $f_{\zeta<0.05}$ \\
\hline
H05 & 46 & $0.005^{+0.018}_{-0.004}$ & 95\% & $0.002^{+0.008}_{-0.001}$ & 98\% \\
V17 & 19 & $0.10^{+0.33}_{-0.09}$ & 40\% & $0.08^{+0.17}_{-0.07}$ & 42\% \\
I20 & 21 & $0.04^{+0.20}_{-0.04}$ & 55\% & $0.04^{+0.18}_{-0.04}$ & 54\% \\

\hline

\hline

\end{tabular}
\tablenotetext{a}{\raggedright The reference velocity distributions of young pulsars, each derived from a velocity sample of sample size $N_\mathrm{s}$, are provided by \citet{Hobbs05}, \citet{Verbunt17}, and \citet{Igoshev20}.}
\tablenotetext{b}{\raggedright $\zeta$ and $f_{\zeta<0.05}$ denote, respectively, the p-value of the null hypothesis (that the magnetar $v_\perp$ sample follows the reference distribution), and the fraction of 10,000 simulations resulting in $\zeta<0.05$.}

\end{table}

Additionally, the magnetar $v_\perp$ sample can, in principle, be used to probe the modality of the underlying $v_\perp$ distribution with, e.g., Hartigan's dip test \citep{Hartigan85,Hartigan85a}.
Following the Monte-Carlo procedure described in Section~7.2.3 of \citet{Ding24}, we obtain the mean dip-test p-value of 0.7, indicating no multimodality of the underlying $v_\perp$ distribution.
When assuming magnetar velocities follow the V17 and I20 young pulsar distributions, we find out with simulation that, respectively, $\sim1590$ and $\sim130$ additional magnetar $v_\perp$ determinations are required to rule out unimodality at 90\% confidence.

Finally, it is important to note that the current magnetar $v_\perp$ sample might be biased against low-velocity magnetars due to relatively low spatial resolution of  optical/infrared facilities. This bias will be reduced with increasing time baseline of astrometry. Additionally, the magnetar sample is biased towards brighter magnetars that tend to be closer ($\lesssim10$\,kpc) to us, though its impact on $v_\perp$ is likely limited.

\section{Summary and Future prospects}
\label{sec:summary_and_prospects}

In this work, we observed the fastest-spinning magnetar \swift\ from April 2020 to March 2023 with the VLBA, in order to determine its astrometric information. To suppress propagation-related systematic errors, the 1D interpolation strategy was adopted. Using a Bayesian framework, we derived (from the measured position series) precise proper motion and parallax for \swift, along with two nuisance parameters that correct the systematic errors.
The parallax $\varpi=0.12\pm0.02$\,mas is the second magnetar parallax, and is among the smallest pulsar parallaxes ever measured; it corresponds to the trigonometric distance $d=9.4^{+2.0}_{-1.6}$\,kpc, locating \swift\ at the far side of the Galactic central region. 
The parallax determination demonstrates that robust and precise trigonometric distance up to $\sim10$\,kpc can be determined for Galactic magnetars, which holds the key to substantially reducing distance-related errors (including systematic errors) in $v_\perp$.

With the trigonometric distance of \swift, the upper limit to the quiescent X-ray luminosity reported by \citet{Esposito20} is refined to $2.1\times10^{34}$\,\ergps, which is more consistent with expectation \citep{Vigano13}.
Combining the pulse broadening $\tau_\mathrm{sc}$ measured by \citet{Lower20a}, the trigonometric distance, and the angular broadening of \swift, we calculate the upper limit $2.6^{+1.8}_{-0.9}$\,kpc of our distance to the closest scattering IISM screen in front of \swift.
In addition, the new magnetar distance agrees with the indicative distance constraint $\gtrsim8$\,kpc of a tentative SNR around \swift, thus sustaining the interpretation that the semi-circular radio feature discovered around \swift\ is an SNR associated with \swift\ \citep{Ibrahim23}.

Incorporating the proper motion of \swift\ with $d$, we obtain a low transverse space velocity $v_\perp=48^{+50}_{-16}$\,\kmps.
With the enriched $v_\perp$ sample of 8 astrometrically studied magnetars, we find that the magnetar $v_\perp$ sample is unlikely to reconcile with the unimodal young pulsar velocity distribution reported by \citet{Hobbs05}, while being marginally consistent with the more recent bimodal young pulsar velocity distributions derived from VLBI results \citep{Verbunt17,Igoshev20}.

Looking into the future, multi-wavelength high-sensitivity observations towards the sky region around \swift\ will constrain the amount of matter blown into the space at the birth of \swift, hence probing the formation channel of \swift. To further refine the magnetar $v_\perp$ distribution, more astrometric measurements of magnetars are desired, which is ultimately limited by the number of identified magnetars in the Galaxy.

\section*{Acknowledgements}
We appreciate the useful comments from the anonymous reviewer, which have well improved the quality of this paper. 
HD acknowledges the EACOA Fellowship awarded by the East Asia Core Observatories Association.
This work is mainly based on observations with the Very Long Baseline Array (VLBA), which is operated by the National Radio Astronomy Observatory (NRAO). 
The NRAO is a facility of the National Science Foundation operated under cooperative agreement by Associated Universities, Inc.
This work made use of the Swinburne University of Technology software correlator, developed as part of the Australian Major National Research Facilities Programme and operated under license.
HD thanks the NRAO VLBA observational support staff (Paul and Anthony) for generous efforts in support of pulsar gating at correlation, including multiple re-correlation passes when required. 
Murriyang, the Parkes radio telescope, is part of the Australia Telescope National Facility (\url{https://ror.org/05qajvd42}) which is funded by the Australian Government for operation as a National Facility managed by CSIRO. We acknowledge the Wiradjuri people as the Traditional Owners of the Observatory site.


\vspace{5mm}
\facilities{Very Long Baseline Array, Murriyang/Parkes radio telescope}

\software{{\tt astropy} \citep{Astropy-Collaboration13,Astropy-Collaboration18,Astropy-Collaboration22}, {\tt numpy} \citep{Harris20}, {\tt scipy} \citep{Virtanen20}, {\tt matplotlib} \citep{Hunter07}, {\tt sterne.py} \citep{Ding21a}, {\tt corner.py} \citep{Foreman-Mackey16}, {\tt bilby} \citep{Ashton19}, {\tt ParselTongue} \citep{Kettenis06}, {\tt AIPS} \citep{Greisen03}, {\tt DIFMAP} \citep{Shepherd94}, {\tt corner.py} \citep{Foreman-Mackey16}, {\tt pygedm} \citep{Price21}, {\tt pmpar}\footnote{\label{footnote:pmpar}\url{https://github.com/walterfb/pmpar}}, {\tt psrqpy} \citep{Pitkin18}}

\bibliography{refs}{}

\begin{thebibliography}{}
\expandafter\ifx\csname natexlab\endcsname\relax\def\natexlab#1{#1}\fi
\providecommand{\url}[1]{\href{#1}{#1}}
\providecommand{\dodoi}[1]{doi:~\href{http://doi.org/#1}{\nolinkurl{#1}}}
\providecommand{\doeprint}[1]{\href{http://ascl.net/#1}{\nolinkurl{http://ascl.net/#1}}}
\providecommand{\doarXiv}[1]{\href{https://arxiv.org/abs/#1}{\nolinkurl{https://arxiv.org/abs/#1}}}

\bibitem[{Ai \& Zhang(2021)}]{Ai21}
Ai, S., \& Zhang, B. 2021, \apjl, 915, L11

\bibitem[{Andersen {et~al.}(2020)Andersen, Bandura, Bhardwaj, Bij, Boyce, Boyle, Brar, Cassanelli, Chawla, Chen, Cliche, Cook, Cubranic, Curtin, Denman, Dobbs, Dong, Fandino, Fonseca, Gaensler, Giri, Good, Halpern, Hill, Hinshaw, H{\"o}fer, Josephy, Kania, Kaspi, Landecker, Leung, Li, Lin, Masui, Mckinven, Mena-Parra, Merryfield, Meyers, Michilli, Milutinovic, Mirhosseini, M{\"u}nchmeyer, Naidu, Newburgh, Ng, Patel, Pen, Pinsonneault-Marotte, Pleunis, Quine, Rafiei-Ravandi, Rahman, Ransom, Renard, Sanghavi, Scholz, Shaw, Shin, Siegel, Singh, Smegal, Smith, Stairs, Tan, Tendulkar, Tretyakov, Vanderlinde, Wang, Wulf, Zwaniga, \& Collaboration}]{Andersen20}
Andersen, B.~C., Bandura, K.~M., Bhardwaj, M., {et~al.} 2020, Nature, 587, 54, \dodoi{10.1038/s41586-020-2863-y}

\bibitem[{Ashton {et~al.}(2019)Ashton, H{\"u}bner, Lasky, Talbot, Ackley, Biscoveanu, Chu, Divakarla, Easter, Goncharov, {et~al.}}]{Ashton19}
Ashton, G., H{\"u}bner, M., Lasky, P.~D., {et~al.} 2019, \apjs, 241, 27, \dodoi{10.3847/1538-4365/ab06fc}

\bibitem[{{Astropy Collaboration} {et~al.}(2013){Astropy Collaboration}, {Robitaille}, {Tollerud}, {Greenfield}, {Droettboom}, {Bray}, {Aldcroft}, {Davis}, {Ginsburg}, {Price-Whelan}, {Kerzendorf}, {Conley}, {Crighton}, {Barbary}, {Muna}, {Ferguson}, {Grollier}, {Parikh}, {Nair}, {Unther}, {Deil}, {Woillez}, {Conseil}, {Kramer}, {Turner}, {Singer}, {Fox}, {Weaver}, {Zabalza}, {Edwards}, {Azalee Bostroem}, {Burke}, {Casey}, {Crawford}, {Dencheva}, {Ely}, {Jenness}, {Labrie}, {Lim}, {Pierfederici}, {Pontzen}, {Ptak}, {Refsdal}, {Servillat}, \& {Streicher}}]{Astropy-Collaboration13}
{Astropy Collaboration}, {Robitaille}, T.~P., {Tollerud}, E.~J., {et~al.} 2013, \aap, 558, A33, \dodoi{10.1051/0004-6361/201322068}

\bibitem[{{Astropy Collaboration} {et~al.}(2018){Astropy Collaboration}, {Price-Whelan}, {Sip{\H{o}}cz}, {G{\"u}nther}, {Lim}, {Crawford}, {Conseil}, {Shupe}, {Craig}, {Dencheva}, {Ginsburg}, {Vand erPlas}, {Bradley}, {P{\'e}rez-Su{\'a}rez}, {de Val-Borro}, {Aldcroft}, {Cruz}, {Robitaille}, {Tollerud}, {Ardelean}, {Babej}, {Bach}, {Bachetti}, {Bakanov}, {Bamford}, {Barentsen}, {Barmby}, {Baumbach}, {Berry}, {Biscani}, {Boquien}, {Bostroem}, {Bouma}, {Brammer}, {Bray}, {Breytenbach}, {Buddelmeijer}, {Burke}, {Calderone}, {Cano Rodr{\'\i}guez}, {Cara}, {Cardoso}, {Cheedella}, {Copin}, {Corrales}, {Crichton}, {D'Avella}, {Deil}, {Depagne}, {Dietrich}, {Donath}, {Droettboom}, {Earl}, {Erben}, {Fabbro}, {Ferreira}, {Finethy}, {Fox}, {Garrison}, {Gibbons}, {Goldstein}, {Gommers}, {Greco}, {Greenfield}, {Groener}, {Grollier}, {Hagen}, {Hirst}, {Homeier}, {Horton}, {Hosseinzadeh}, {Hu}, {Hunkeler}, {Ivezi{\'c}}, {Jain}, {Jenness}, {Kanarek}, {Kendrew}, {Kern}, {Kerzendorf}, {Khvalko}, {King}, {Kirkby}, {Kulkarni},
  {Kumar}, {Lee}, {Lenz}, {Littlefair}, {Ma}, {Macleod}, {Mastropietro}, {McCully}, {Montagnac}, {Morris}, {Mueller}, {Mumford}, {Muna}, {Murphy}, {Nelson}, {Nguyen}, {Ninan}, {N{\"o}the}, {Ogaz}, {Oh}, {Parejko}, {Parley}, {Pascual}, {Patil}, {Patil}, {Plunkett}, {Prochaska}, {Rastogi}, {Reddy Janga}, {Sabater}, {Sakurikar}, {Seifert}, {Sherbert}, {Sherwood-Taylor}, {Shih}, {Sick}, {Silbiger}, {Singanamalla}, {Singer}, {Sladen}, {Sooley}, {Sornarajah}, {Streicher}, {Teuben}, {Thomas}, {Tremblay}, {Turner}, {Terr{\'o}n}, {van Kerkwijk}, {de la Vega}, {Watkins}, {Weaver}, {Whitmore}, {Woillez}, {Zabalza}, \& {Astropy Contributors}}]{Astropy-Collaboration18}
{Astropy Collaboration}, {Price-Whelan}, A.~M., {Sip{\H{o}}cz}, B.~M., {et~al.} 2018, \aj, 156, 123, \dodoi{10.3847/1538-3881/aabc4f}

\bibitem[{{Astropy Collaboration} {et~al.}(2022){Astropy Collaboration}, {Price-Whelan}, {Lim}, {Earl}, {Starkman}, {Bradley}, {Shupe}, {Patil}, {Corrales}, {Brasseur}, {N{"o}the}, {Donath}, {Tollerud}, {Morris}, {Ginsburg}, {Vaher}, {Weaver}, {Tocknell}, {Jamieson}, {van Kerkwijk}, {Robitaille}, {Merry}, {Bachetti}, {G{"u}nther}, {Aldcroft}, {Alvarado-Montes}, {Archibald}, {B{'o}di}, {Bapat}, {Barentsen}, {Baz{'a}n}, {Biswas}, {Boquien}, {Burke}, {Cara}, {Cara}, {Conroy}, {Conseil}, {Craig}, {Cross}, {Cruz}, {D'Eugenio}, {Dencheva}, {Devillepoix}, {Dietrich}, {Eigenbrot}, {Erben}, {Ferreira}, {Foreman-Mackey}, {Fox}, {Freij}, {Garg}, {Geda}, {Glattly}, {Gondhalekar}, {Gordon}, {Grant}, {Greenfield}, {Groener}, {Guest}, {Gurovich}, {Handberg}, {Hart}, {Hatfield-Dodds}, {Homeier}, {Hosseinzadeh}, {Jenness}, {Jones}, {Joseph}, {Kalmbach}, {Karamehmetoglu}, {Ka{l}uszy{'n}ski}, {Kelley}, {Kern}, {Kerzendorf}, {Koch}, {Kulumani}, {Lee}, {Ly}, {Ma}, {MacBride}, {Maljaars}, {Muna}, {Murphy}, {Norman}, {O'Steen},
  {Oman}, {Pacifici}, {Pascual}, {Pascual-Granado}, {Patil}, {Perren}, {Pickering}, {Rastogi}, {Roulston}, {Ryan}, {Rykoff}, {Sabater}, {Sakurikar}, {Salgado}, {Sanghi}, {Saunders}, {Savchenko}, {Schwardt}, {Seifert-Eckert}, {Shih}, {Jain}, {Shukla}, {Sick}, {Simpson}, {Singanamalla}, {Singer}, {Singhal}, {Sinha}, {Sip{H{o}}cz}, {Spitler}, {Stansby}, {Streicher}, {{{S}}umak}, {Swinbank}, {Taranu}, {Tewary}, {Tremblay}, {Val-Borro}, {Van Kooten}, {Vasovi{'c}}, {Verma}, {de Miranda Cardoso}, {Williams}, {Wilson}, {Winkel}, {Wood-Vasey}, {Xue}, {Yoachim}, {Zhang}, {Zonca}, \& {Astropy Project Contributors}}]{Astropy-Collaboration22}
{Astropy Collaboration}, {Price-Whelan}, A.~M., {Lim}, P.~L., {et~al.} 2022, \apj, 935, 167, \dodoi{10.3847/1538-4357/ac7c74}

\bibitem[{Bailer-Jones {et~al.}(2021)Bailer-Jones, Rybizki, Fouesneau, Demleitner, \& Andrae}]{Bailer-Jones21}
Bailer-Jones, C., Rybizki, J., Fouesneau, M., Demleitner, M., \& Andrae, R. 2021, \aj, 161, 147, \dodoi{10.3847/1538-3881/abd806}

\bibitem[{Bailes {et~al.}(2021)Bailes, Bassa, Bernardi, Buchner, Burgay, Caleb, Cooper, Desvignes, Groot, Heywood, {et~al.}}]{Bailes21}
Bailes, M., Bassa, C., Bernardi, G., {et~al.} 2021, \mnras, 503, 5367

\bibitem[{Benli \& Ertan(2016)}]{Benli16}
Benli, O., \& Ertan, {\"U}. 2016, \mnras, 457, 4114

\bibitem[{Bochenek {et~al.}(2020)Bochenek, Ravi, Belov, Hallinan, Kocz, Kulkarni, \& McKenna}]{Bochenek20}
Bochenek, C.~D., Ravi, V., Belov, K.~V., {et~al.} 2020, Nature, 587, 59

\bibitem[{Borkowski \& Reynolds(2017)}]{Borkowski17}
Borkowski, K.~J., \& Reynolds, S.~P. 2017, \apj, 846, 13

\bibitem[{Bower {et~al.}(2014)Bower, Deller, Demorest, Brunthaler, Eatough, Falcke, Kramer, Lee, \& Spitler}]{Bower14}
Bower, G.~C., Deller, A., Demorest, P., {et~al.} 2014, \apjl, 780, L2

\bibitem[{Bower {et~al.}(2015)Bower, Deller, Demorest, Brunthaler, Falcke, Moscibrodzka, O'Leary, Eatough, Kramer, Lee, {et~al.}}]{Bower15}
---. 2015, \apj, 798, 120

\bibitem[{Camilo {et~al.}(2008)Camilo, Reynolds, Johnston, Halpern, \& Ransom}]{Camilo08}
Camilo, F., Reynolds, J., Johnston, S., Halpern, J., \& Ransom, S. 2008, \apj, 679, 681

\bibitem[{Camilo {et~al.}(2007)Camilo, Cognard, Ransom, Halpern, Reynolds, Zimmerman, Gotthelf, Helfand, Demorest, Theureau, {et~al.}}]{Camilo07}
Camilo, F., Cognard, I., Ransom, S., {et~al.} 2007, \apj, 663, 497

\bibitem[{Champion {et~al.}(2020)Champion, Cognard, Cruces, Desvignes, Jankowski, Karuppusamy, Keith, Kouveliotou, Kramer, Liu, {et~al.}}]{Champion20a}
Champion, D., Cognard, I., Cruces, M., {et~al.} 2020, \mnras, 498, 6044

\bibitem[{Cordes {et~al.}(1983)Cordes, Weisberg, \& Boriakoff}]{Cordes83}
Cordes, J., Weisberg, J., \& Boriakoff, V. 1983, \apj, 268, 370

\bibitem[{Cordes {et~al.}(1985)Cordes, Weisberg, \& Boriakoff}]{Cordes85}
---. 1985, \apj, 288, 221

\bibitem[{Cordes(2005)}]{Cordes05}
Cordes, J.~M. 2005, in Low Frequency Astrophysics from Space: Proceedings of an International Workshop Held in Crystal City, Virginia, USA, on 8 and 9 January 1990, Springer, 165--174

\bibitem[{Cordes \& Lazio(1991)}]{Cordes91}
Cordes, J.~M., \& Lazio, T.~J. 1991, \apj, 376, 123

\bibitem[{Cordes \& Lazio(2002)}]{Cordes02}
Cordes, J.~M., \& Lazio, T. J.~W. 2002, arXiv preprint astro-ph/0207156

\bibitem[{Deller {et~al.}(2012)Deller, Camilo, Reynolds, \& Halpern}]{Deller12a}
Deller, A., Camilo, F., Reynolds, J., \& Halpern, J. 2012, \apjl, 748, L1

\bibitem[{{Deller} {et~al.}(2011){Deller}, {Brisken}, {Phillips}, {Morgan}, {Alef}, {Cappallo}, {Middelberg}, {Romney}, {Rottmann}, {Tingay}, \& {Wayth}}]{Deller11a}
{Deller}, A.~T., {Brisken}, W.~F., {Phillips}, C.~J., {et~al.} 2011, \pasp, 123, 275, \dodoi{10.1086/658907}

\bibitem[{{Deller} {et~al.}(2019){Deller}, {Goss}, {Brisken}, {Chatterjee}, {Cordes}, {Janssen}, {Kovalev}, {Lazio}, {Petrov}, {Stappers}, \& {Lyne}}]{Deller19}
{Deller}, A.~T., {Goss}, W.~M., {Brisken}, W.~F., {et~al.} 2019, \apj, 875, 100, \dodoi{10.3847/1538-4357/ab11c7}

\bibitem[{Dessart {et~al.}(2007)Dessart, Burrows, Livne, \& Ott}]{Dessart07}
Dessart, L., Burrows, A., Livne, E., \& Ott, C. 2007, \apj, 669, 585

\bibitem[{Ding(2022)}]{Ding22}
Ding, H. 2022, PhD thesis, Swinburne University of Technology.
\newblock \url{https://arxiv.org/abs/2212.08881}

\bibitem[{Ding \& Deller(2024)}]{Ding24a}
Ding, H., \& Deller, A. 2024, dingswin/sterne: v2.1.1,  Zenodo, \dodoi{10.5281/ZENODO.11239560}

\bibitem[{Ding {et~al.}(2023{\natexlab{a}})Ding, Deller, Lower, \& Shannon}]{Ding23a}
Ding, H., Deller, A., Lower, M., \& Shannon, R. 2023{\natexlab{a}}, Proceedings of the International Astronomical Union, 16, 271, \dodoi{10.1017/S1743921322000321}

\bibitem[{Ding {et~al.}(2021)Ding, Deller, Fonseca, Stairs, Stappers, \& Lyne}]{Ding21a}
Ding, H., Deller, A.~T., Fonseca, E., {et~al.} 2021, \apjl, 921, L19, \dodoi{10.3847/2041-8213/ac3091}

\bibitem[{Ding {et~al.}(2020{\natexlab{a}})Ding, Deller, Freire, Kaplan, Lazio, Shannon, \& Stappers}]{Ding20}
Ding, H., Deller, A.~T., Freire, P., {et~al.} 2020{\natexlab{a}}, \apj, 896, 85, \dodoi{10.3847/1538-4357/ab8f27}

\bibitem[{Ding {et~al.}(2020{\natexlab{b}})Ding, Deller, Lower, \& Shannon}]{Ding20b}
Ding, H., Deller, A.~T., Lower, M.~E., \& Shannon, R.~M. 2020{\natexlab{b}}, ATel, 14005, 1

\bibitem[{Ding {et~al.}(2024)Ding, Deller, Swiggum, Lynch, Chatterjee, \& Tauris}]{Ding24}
Ding, H., Deller, A.~T., Swiggum, J.~K., {et~al.} 2024, arXiv preprint arXiv:2405.03914

\bibitem[{Ding {et~al.}(2020{\natexlab{c}})Ding, Deller, Lower, Flynn, Chatterjee, Brisken, Hurley-Walker, Camilo, Sarkissian, \& Gupta}]{Ding20c}
Ding, H., Deller, A.~T., Lower, M.~E., {et~al.} 2020{\natexlab{c}}, \mnras, 498, 3736, \dodoi{10.1093/mnras/staa2531}

\bibitem[{Ding {et~al.}(2023{\natexlab{b}})Ding, Deller, Stappers, Lazio, Kaplan, Chatterjee, Brisken, Cordes, Freire, Fonseca, Stairs, Guillemot, Lyne, Cognard, Reardon, \& Theureau}]{Ding23}
Ding, H., Deller, A.~T., Stappers, B.~W., {et~al.} 2023{\natexlab{b}}, \mnras, 519, 4982, \dodoi{10.1093/mnras/stac3725}

\bibitem[{Doi {et~al.}(2006)Doi, Fujisawa, Habe, Honma, Kawaguchi, Kobayashi, Murata, Omodaka, Sudou, \& Takaba}]{Doi06}
Doi, A., Fujisawa, K., Habe, A., {et~al.} 2006, \pasj, 58, 777

\bibitem[{Efron \& Tibshirani(1994)}]{Efron94}
Efron, B., \& Tibshirani, R.~J. 1994, An introduction to the bootstrap (Chapman and Hall/CRC)

\bibitem[{Enoto {et~al.}(2020)Enoto, Sakamoto, Younes, Hu, Ho, Gendreau, Arzoumanian, Guver, Guillot, Altamirano, {et~al.}}]{Enoto20}
Enoto, T., Sakamoto, T., Younes, G., {et~al.} 2020, ATel, 13551, 1

\bibitem[{Esposito {et~al.}(2020)Esposito, Rea, Borghese, Zelati, Vigan{\`o}, Israel, Tiengo, Ridolfi, Possenti, Burgay, {et~al.}}]{Esposito20}
Esposito, P., Rea, N., Borghese, A., {et~al.} 2020, \apjl, 896, L30

\bibitem[{Fomalont \& Kopeikin(2003)}]{Fomalont03}
Fomalont, E.~B., \& Kopeikin, S.~M. 2003, \apj, 598, 704

\bibitem[{Foreman-Mackey(2016)}]{Foreman-Mackey16}
Foreman-Mackey, D. 2016, The Journal of Open Source Software, 1, 24, \dodoi{10.21105/joss.00024}

\bibitem[{Fryer {et~al.}(1999)Fryer, Benz, Herant, \& Colgate}]{Fryer99}
Fryer, C., Benz, W., Herant, M., \& Colgate, S.~A. 1999, \apj, 516, 892

\bibitem[{Gaensler(2014)}]{Gaensler14}
Gaensler, B. 2014, GRB Coordinates Network, 16533, 1

\bibitem[{Gaensler \& Chatterjee(2008)}]{Gaensler08}
Gaensler, B., \& Chatterjee, S. 2008, GRB Coordinates Network, 8149, 1

\bibitem[{Gelfand \& Gaensler(2007)}]{Gelfand07}
Gelfand, J.~D., \& Gaensler, B. 2007, \apj, 667, 1111

\bibitem[{Giacomazzo \& Perna(2013)}]{Giacomazzo13}
Giacomazzo, B., \& Perna, R. 2013, \apjl, 771, L26

\bibitem[{Goodman \& Narayan(1989)}]{Goodman89}
Goodman, J., \& Narayan, R. 1989, \mnras, 238, 995

\bibitem[{{Greisen}(2003)}]{Greisen03}
{Greisen}, E.~W. 2003, in Astrophysics and Space Science Library, Vol. 285, Information Handling in Astronomy - Historical Vistas, ed. A.~{Heck} (Springer), 109, \dodoi{10.1007/0-306-48080-8_7}

\bibitem[{Harris {et~al.}(2020)Harris, Millman, Van Der~Walt, Gommers, Virtanen, Cournapeau, Wieser, Taylor, Berg, Smith, {et~al.}}]{Harris20}
Harris, C.~R., Millman, K.~J., Van Der~Walt, S.~J., {et~al.} 2020, Nature, 585, 357

\bibitem[{Hartigan \& Hartigan(1985)}]{Hartigan85}
Hartigan, J.~A., \& Hartigan, P.~M. 1985, The annals of Statistics, 70

\bibitem[{Hartigan(1985)}]{Hartigan85a}
Hartigan, P. 1985, Journal of the Royal Statistical Society. Series C (Applied Statistics), 34, 320

\bibitem[{{Helfand} {et~al.}(2007){Helfand}, {Chatterjee}, {Brisken}, {Camilo}, {Reynolds}, {van Kerkwijk}, {Halpern}, \& {Ransom}}]{Helfand07}
{Helfand}, D.~J., {Chatterjee}, S., {Brisken}, W.~F., {et~al.} 2007, \apj, 662, 1198, \dodoi{10.1086/518028}

\bibitem[{Heyl \& Kulkarni(1998)}]{Heyl98}
Heyl, J.~S., \& Kulkarni, S. 1998, \apj, 506, L61

\bibitem[{Hobbs {et~al.}(2005)Hobbs, Lorimer, Lyne, \& Kramer}]{Hobbs05}
Hobbs, G., Lorimer, D., Lyne, A., \& Kramer, M. 2005, \mnras, 360, 974

\bibitem[{Hunter(2007)}]{Hunter07}
Hunter, J.~D. 2007, Computing in science \& engineering, 9, 90

\bibitem[{Ibrahim {et~al.}(2023)Ibrahim, Borghese, Rea, Zelati, Parent, Russell, Ascenzi, Sathyaprakash, G{\"o}tz, Mereghetti, {et~al.}}]{Ibrahim23}
Ibrahim, A., Borghese, A., Rea, N., {et~al.} 2023, \apj, 943, 20

\bibitem[{Igoshev {et~al.}(2016)Igoshev, Verbunt, \& Cator}]{Igoshev16}
Igoshev, A., Verbunt, F., \& Cator, E. 2016, Astronomy \& Astrophysics, 591, A123

\bibitem[{Igoshev(2020)}]{Igoshev20}
Igoshev, A.~P. 2020, \mnras, 494, 3663

\bibitem[{Kaplan {et~al.}(2008)Kaplan, Chatterjee, Hales, Gaensler, \& Slane}]{Kaplan08}
Kaplan, D., Chatterjee, S., Hales, C., Gaensler, B., \& Slane, P. 2008, \apj, 137, 354

\bibitem[{Keane \& Kramer(2008)}]{Keane08}
Keane, E.~F., \& Kramer, M. 2008, \mnras, 391, 2009

\bibitem[{{Kettenis} {et~al.}(2006){Kettenis}, {van Langevelde}, {Reynolds}, \& {Cotton}}]{Kettenis06}
{Kettenis}, M., {van Langevelde}, H.~J., {Reynolds}, C., \& {Cotton}, B. 2006, in Astronomical Society of the Pacific Conference Series, Vol. 351, Astronomical Data Analysis Software and Systems XV, ed. C.~{Gabriel}, C.~{Arviset}, D.~{Ponz}, \& S.~{Enrique}, 497

\bibitem[{Klose {et~al.}(2004)Klose, Henden, Geppert, Greiner, Guetter, Hartmann, Kouveliotou, Luginbuhl, Stecklum, \& Vrba}]{Klose04}
Klose, S., Henden, A., Geppert, U., {et~al.} 2004, \apj, 609, L13

\bibitem[{Lipunov \& Postnov(1985)}]{Lipunov85}
Lipunov, V., \& Postnov, K. 1985, \aap, 144, L13

\bibitem[{Lorimer {et~al.}(2006)Lorimer, Faulkner, Lyne, Manchester, Kramer, McLaughlin, Hobbs, Possenti, Stairs, Camilo, {et~al.}}]{Lorimer06a}
Lorimer, D., Faulkner, A., Lyne, A., {et~al.} 2006, \mnras, 372, 777

\bibitem[{Lower {et~al.}(2020{\natexlab{a}})Lower, Johnston, Shannon, Bailes, \& Camilo}]{Lower20c}
Lower, M.~E., Johnston, S., Shannon, R.~M., Bailes, M., \& Camilo, F. 2020{\natexlab{a}}, \mnras

\bibitem[{Lower {et~al.}(2020{\natexlab{b}})Lower, Shannon, Johnston, \& Bailes}]{Lower20a}
Lower, M.~E., Shannon, R.~M., Johnston, S., \& Bailes, M. 2020{\natexlab{b}}, \apjl, 896, L37

\bibitem[{{Lutz} \& {Kelker}(1973)}]{Lutz73}
{Lutz}, T.~E., \& {Kelker}, D.~H. 1973, \pasp, 85, 573, \dodoi{10.1086/129506}

\bibitem[{Lyman {et~al.}(2022)Lyman, Levan, Wiersema, Kouveliotou, Chrimes, \& Fruchter}]{Lyman22}
Lyman, J., Levan, A., Wiersema, K., {et~al.} 2022, \apj, 926, 121

\bibitem[{Macquart \& Koay(2013)}]{Macquart13}
Macquart, J.-P., \& Koay, J.~Y. 2013, \apj, 776, 125

\bibitem[{Mannings {et~al.}(2021)Mannings, Fong, Simha, Prochaska, Rafelski, Kilpatrick, Tejos, Heintz, Bannister, Bhandari, {et~al.}}]{Mannings21}
Mannings, A.~G., Fong, W.-f., Simha, S., {et~al.} 2021, \apj, 917, 75

\bibitem[{Margalit {et~al.}(2019)Margalit, Berger, \& Metzger}]{Margalit19}
Margalit, B., Berger, E., \& Metzger, B.~D. 2019, \apj, 886, 110

\bibitem[{Ocker {et~al.}(2022)Ocker, Cordes, Chatterjee, \& Gorsuch}]{Ocker22}
Ocker, S.~K., Cordes, J.~M., Chatterjee, S., \& Gorsuch, M.~R. 2022, \apj, 934, 71

\bibitem[{{Olausen} \& {Kaspi}(2014)}]{Olausen14}
{Olausen}, S.~A., \& {Kaspi}, V.~M. 2014, \apjs, 212, 6, \dodoi{10.1088/0067-0049/212/1/6}

\bibitem[{Oswald {et~al.}(2021)Oswald, Karastergiou, Posselt, Johnston, Bailes, Buchner, Geyer, Keith, Kramer, Parthasarathy, {et~al.}}]{Oswald21}
Oswald, L., Karastergiou, A., Posselt, B., {et~al.} 2021, \mnras, 504, 1115

\bibitem[{Pitkin(2018)}]{Pitkin18}
Pitkin, M. 2018, arXiv preprint arXiv:1806.07809

\bibitem[{Price {et~al.}(2021)Price, Flynn, \& Deller}]{Price21}
Price, D.~C., Flynn, C., \& Deller, A. 2021, \pasa, 38, e038

\bibitem[{Rajwade {et~al.}(2022)Rajwade, Stappers, Lyne, Shaw, Mickaliger, Liu, Kramer, Desvignes, Karuppusamy, Enoto, {et~al.}}]{Rajwade22}
Rajwade, K., Stappers, B., Lyne, A., {et~al.} 2022, \mnras, 512, 1687

\bibitem[{Rea {et~al.}(2012)Rea, Pons, Torres, \& Turolla}]{Rea12}
Rea, N., Pons, J.~A., Torres, D.~F., \& Turolla, R. 2012, \apjl, 748, L12

\bibitem[{Ruiter {et~al.}(2019)Ruiter, Ferrario, Belczynski, Seitenzahl, Crocker, \& Karakas}]{Ruiter19}
Ruiter, A., Ferrario, L., Belczynski, K., {et~al.} 2019, \mnras, 484, 698

\bibitem[{Schneider {et~al.}(2019)Schneider, Ohlmann, Podsiadlowski, R{\"o}pke, Balbus, Pakmor, \& Springel}]{Schneider19}
Schneider, F.~R., Ohlmann, S.~T., Podsiadlowski, P., {et~al.} 2019, Nature, 574, 211

\bibitem[{Sedov(2018)}]{Sedov18}
Sedov, L.~I. 2018, Similarity and dimensional methods in mechanics (CRC press)

\bibitem[{Shenar {et~al.}(2023)Shenar, Wade, Marchant, Bagnulo, Bodensteiner, Bowman, Gilkis, Langer, Nicolas-Chen{\'e}, Oskinova, {et~al.}}]{Shenar23}
Shenar, T., Wade, G.~A., Marchant, P., {et~al.} 2023, Science, 381, 761

\bibitem[{{Shepherd} {et~al.}(1994){Shepherd}, {Pearson}, \& {Taylor}}]{Shepherd94}
{Shepherd}, M.~C., {Pearson}, T.~J., \& {Taylor}, G.~B. 1994, in \baas, Vol.~26, Bulletin of the American Astronomical Society, 987--989

\bibitem[{Sherman {et~al.}(2024)Sherman, Ravi, El-Badry, Sharma, Ocker, Kosogorov, Connor, \& Faber}]{Sherman24}
Sherman, M.~B., Ravi, V., El-Badry, K., {et~al.} 2024, \mnras, stae1289

\bibitem[{{Sokolovsky} {et~al.}(2011){Sokolovsky}, {Kovalev}, {Pushkarev}, \& {Lobanov}}]{Sokolovsky11}
{Sokolovsky}, K.~V., {Kovalev}, Y.~Y., {Pushkarev}, A.~B., \& {Lobanov}, A.~P. 2011, \aap, 532, A38, \dodoi{10.1051/0004-6361/201016072}

\bibitem[{Sun {et~al.}(2019)Sun, Li, Zhang, Zhang, Bauer, Xue, \& Yuan}]{Sun19}
Sun, H., Li, Y., Zhang, B.-B., {et~al.} 2019, \apj, 886, 129

\bibitem[{Tendulkar {et~al.}(2012)Tendulkar, Cameron, \& Kulkarni}]{Tendulkar12}
Tendulkar, S.~P., Cameron, P.~B., \& Kulkarni, S.~R. 2012, \apj, 761, 76

\bibitem[{Tendulkar {et~al.}(2013)Tendulkar, Cameron, \& Kulkarni}]{Tendulkar13}
---. 2013, \apj, 772, 31

\bibitem[{Thompson \& Duncan(1995)}]{Thompson95}
Thompson, C., \& Duncan, R.~C. 1995, \mnras, 275, 255

\bibitem[{Torne {et~al.}(2020)Torne, Liu, Cognard, Desvignes, Karuppusamy, Kramer, Paubert, Lyne, Rajwade, Stappers, {et~al.}}]{Torne20}
Torne, P., Liu, K., Cognard, I., {et~al.} 2020, ATel, 14001, 1

\bibitem[{Truelove \& McKee(1999)}]{Truelove99}
Truelove, J.~K., \& McKee, C.~F. 1999, \apjs, 120, 299

\bibitem[{Vasisht \& Gotthelf(1997)}]{Vasisht97}
Vasisht, G., \& Gotthelf, E. 1997, \apjl, 486, L129

\bibitem[{{Verbunt} {et~al.}(2017){Verbunt}, {Igoshev}, \& {Cator}}]{Verbunt17}
{Verbunt}, F., {Igoshev}, A., \& {Cator}, E. 2017, \aap, 608, A57, \dodoi{10.1051/0004-6361/201731518}

\bibitem[{Vigan{\`o} {et~al.}(2013)Vigan{\`o}, Rea, Pons, Perna, Aguilera, \& Miralles}]{Vigano13}
Vigan{\`o}, D., Rea, N., Pons, J.~A., {et~al.} 2013, \mnras, 434, 123

\bibitem[{Virtanen {et~al.}(2020)Virtanen, Gommers, Oliphant, Haberland, Reddy, Cournapeau, Burovski, Peterson, Weckesser, Bright, {van der Walt}, Brett, Wilson, Millman, Mayorov, Nelson, Jones, Kern, Larson, Carey, Polat, Feng, Moore, {VanderPlas}, Laxalde, Perktold, Cimrman, Henriksen, Quintero, Harris, Archibald, Ribeiro, Pedregosa, {van Mulbregt}, \& {SciPy 1.0 Contributors}}]{Virtanen20}
Virtanen, P., Gommers, R., Oliphant, T.~E., {et~al.} 2020, Nature Methods, 17, 261, \dodoi{10.1038/s41592-019-0686-2}

\bibitem[{Xue {et~al.}(2019)Xue, Zheng, Li, Brandt, Zhang, Luo, Zhang, Bauer, Sun, Lehmer, {et~al.}}]{Xue19}
Xue, Y., Zheng, X., Li, Y., {et~al.} 2019, Nature, 568, 198

\bibitem[{{Yao} {et~al.}(2017){Yao}, {Manchester}, \& {Wang}}]{Yao17}
{Yao}, J.~M., {Manchester}, R.~N., \& {Wang}, N. 2017, \apj, 835, 29, \dodoi{10.3847/1538-4357/835/1/29}

\end{thebibliography}
\bibliographystyle{aasjournal}

\end{document}